\documentclass{aastex}
\usepackage{emulateapj5}
\usepackage{onecolfloat5}
\usepackage{apjfonts}

\def\plotone#1{\centering \leavevmode
\epsfxsize= 1.0\columnwidth \epsfbox{#1}}

\def\gsim{\;\rlap{\lower 2.5pt
 \hbox{$\sim$}}\raise 1.5pt\hbox{$>$}\;}
\def\lsim{\;\rlap{\lower 2.5pt
   \hbox{$\sim$}}\raise 1.5pt\hbox{$<$}\;}
\newcommand{\be}{\begin{equation}}
\newcommand{\beq}{\begin{equation}}
\newcommand{\ba}{\begin{eqnarray}}
\newcommand{\ee}{\end{equation}}
\newcommand{\eeq}{\end{equation}}
\newcommand{\ea}{\end{eqnarray}}

\newcommand\msun{\rm M_\odot}

\usepackage{natbib}

\newcommand{\cel}{C_{E\ell}}
\newcommand{\ctl}{C_{T\ell}}
\newcommand{\ccl}{C_{C\ell}}

\begin{document}
\twocolumn[
\submitted{Submitted to ApJ}
\title{The Reionization History at High Redshifts I: Physical Models and New Constraints from CMB Polarization}

\author{Zolt\'an Haiman}
\affil{Department of Astronomy, Columbia University, 550 West 120th Street, New York, NY 10027}
\author{Gilbert P. Holder}
\affil{Institute for Advanced Study, School of Natural Sciences, Olden Lane, Princeton, NJ 08540}

\begin{abstract}
The recent discovery of a high optical depth $\tau$ to Thomson
scattering from the Wilkinson Microwave Anisotropy Probe ({\it WMAP})
data implies that significant reionization took place at redshifts
$z>6$.  This discovery has important implications for the sources of
reionization, and allows, for the first time, constraints to be placed
on physical reionization scenarios out to redshift $z\sim 20$.  Using
a new suite of semi--analytic reionization models, we show that the
high value of $\tau$ requires a surprisingly high efficiency
$\epsilon$ of the first generation of UV sources for injecting
ionizing photons into the intergalactic medium. We find that no simple
reionization model can be consistent with the combination of the {\it
WMAP} result with data from the $z\lsim 6.5$ universe.  Satisfying
both constraints requires either of the following: (i) ${\rm H_2}$
molecules form efficiently at $z\sim 20$, survive feedback processes,
and allow UV sources in halos with virial temperatures $T_{\rm
vir}<10^4$K to contribute substantially to reionization, or (ii) the
efficiency $\epsilon$ in halos with $T_{\rm vir}>10^4$K decreased by a
factor of $\gsim 30$ between $z\sim 20$ and $z\sim 6$.  We discuss the
relevant physical issues to produce either scenario, and argue that
both options are viable, and allowed by current data.  In detailed
models of the reionization history, we find that the evolution of the
ionized fractions in the two scenarios have distinctive features that
Planck can distinguish at $\gsim 3\sigma$ significance.  At the high
{\it WMAP} value for $\tau$, Planck will also be able to provide tight
statistical constraints on reionization model parameters, and
elucidate much of the physics at the end of the Dark Ages.  The
sources responsible for the high optical depth discovered by {\it
WMAP} should be directly detectable out to $z\sim 15$ by the {\it
James Webb Space Telescope}.
\end{abstract}]


\section{Introduction}
\label{sec:introduction}
How and when the intergalactic medium (IGM) was reionized is one of
the long outstanding questions in astrophysical cosmology, holding
many clues about the onset of structure formation in cold dark matter
(CDM) cosmologies, and the nature of the first generation of light
sources. The past year has seen an explosion of progress in probing
the reionization history, culminating in recent results from the {\it WMAP}
satellite (Bennett et al. 2003).  Spectroscopic observations of high
redshift quasars had previously found no strong HI absorption (a
so--called ``Gunn--Peterson '', GP trough) in sources at $z \lsim 6$,
showing that the IGM is highly ionized at these redshifts (Fan et
al. 2000).  However, the discoveries of a handful of bright quasars in
the Sloan Digital Sky Survey (SDSS) at redshifts $z>6$ have revealed
full GP troughs, i.e. spectra consistent with no flux at high S/N over
a substantial stretch of wavelength shortward of
$(1+z)\lambda_\alpha=8850$\AA\ (Becker et al. 2001; Fan et al. 2003).
The lack of any detectable flux translates to strong lower limits
$x_{\rm H}\gsim 0.01$ on the mean mass--weighted neutral fraction of
the IGM at $z\sim 6$ (Cen \& McDonald 2002; Fan et al. 2002; Lidz et
al. 2002; Pentericci et al. 2002).  Comparisons with numerical
simulations of cosmological reionization (Gnedin 2001; Cen \& McDonald
2002; Fan et al. 2002), together with the rapid rise towards high
redshifts of the neutral fractions inferred from a sample of high
redshift quasars from $5\lsim z \lsim 6$ (Songaila \& Cowie 2002),
suggest that the IGM becomes neutral at a relatively narrow redshift
interval, $\Delta z\sim 1$, beyond the redshifts of the most distant
sources.

An alternative way of probing the reionization history is to study the
temperature and polarization anisotropies of the cosmic microwave
background (CMB).  Both anisotropy patterns are affected by Thomson
scattering of CMB photons off the free electrons in a reionized IGM
(Zaldarriaga 1997).  Physically, the CMB and the GP troughs probe two
different stages of reionization: the CMB is sensitive to the initial
phase when $x_{\rm H}$ first decreases below unity, and free electrons
appear; say, at redshift $z_{\rm e}$.  On the other hand, the
(hydrogen) GP trough is sensitive to the late stages, when the
remaining neutral hydrogen atoms are cleared away, at redshift $z_{\rm
H}$ (Kaplinghat et al. 2003).  The recent measurement by the {\it
WMAP} (Bennett et al. 2003) satellite of the optical depth,
$\tau=0.17\pm0.04$ to Thomson scattering (Kogut et al. 2003) implies a
reionization redshift of $z_e=17\pm5$, under the assumption that
reionization occurs abruptly.  This is to be compared to $z_{\rm
H}\sim 6-7$, obtained from extrapolations of the available high
redshift quasar spectra at $z\sim 6$. If the redshift $z_{\rm H}$
coincided with $z_e$, it would imply optical depths of only $\tau\sim
0.4-0.5$.  While the formal discrepancy between this opacity and the
{\it WMAP} value is only at the $\sim 3\sigma$ level, it is timely to
ask the following questions: {\em (1) what are the physical
requirements on the ionizing sources in a $\Lambda$CDM cosmology to
produce an optical depth of $\tau\sim0.17$, and (2) are these
consistent with other data at $z\lsim 6$?}

In this paper, we address both of these questions. In the $\Lambda$CDM
cosmology favored by {\it WMAP}, $\sim 3$ sigma peaks of the primordial
density field on mass scales of $M\sim 10^6~{\rm M_\odot}$ collapse at
redshifts of $z\sim 20$.  The onset of reionization at similar
redshifts is therefore dependent on efficient formation of ionizing
sources in nonlinear objects down to these low mass--scales.  Early
reionization can be accommodated in models of reionization in
$\Lambda$CDM cosmologies (Shapiro, Giroux \& Babul 1994; Tegmark et
al. 1994; Haiman \& Loeb 1997, 1998; Valageas \& Silk 1999; Haiman,
Abel \& Rees 2000), and is indeed a natural consequence in models
including a sufficient number of metal--free stars (Wyithe \& Loeb
2003 and Cen 2003). If the optical depth is as high as the central
value determined by {\it WMAP}, then future CMB polarization measurements
can provide significant constraint on such models (Kaplinghat et
al. 2003, see also Venkatesan 2000, 2002).

This paper is organized as follows: in \S~\ref{sec:physics}, we
discuss the relevant physical effects that can lead to reionization
histories consistent with both the {\it WMAP} data and the $z\sim 6$ quasar
spectra . In \S~\ref{sec:models}, we then describe a semi--analytical
model that takes these effects into account, and in
\S~\ref{sec:histories}, we derive and discuss the ionization histories
in a suite of models. In \S~\ref{sec:cmb}, we quantify constraints on
the models that will be available from future polarization
measurements by {\it WMAP} and by the Planck satellite.\footnote{CMB
constraints are discussed in more detail in a companion paper (paper
II, Holder et al. 2003), focusing on information beyond measuring
$\tau$, and on the possible bias in the values of $\tau$ measured
without a--priori knowledge of the reionization history.} Motivated by
the required presence of ionizing sources at $z\sim 20$, in
\S~\ref{sec:jwst} we consider the highest redshift at which the {\it
James Webb Space Telescope} can detect these sources.  Finally, we
discuss our results and offer our conclusions in
\S~\ref{sec:discussion}. In this paper we adopt the background
cosmological parameters as measured by the {\it WMAP} experiment (Spergel et
al. 2003, Tables 1 and 2), $\Omega_m=0.29$, $\Omega_{\Lambda}=0.71$,
$\Omega_b=0.047$, $h=0.72$ and an initial matter power spectrum $P(k)
\propto k^n$ with $n=0.99$ and normalization $\sigma_8=0.9$; with the
exception of \S~\ref{subsec:running}, where we consider the running
index model favored by the combination of {\it WMAP} with other data
(Spergel et al. 2003, Table 10).

\section{Summary of Relevant Physics at High Redshifts}
\label{sec:physics}

In the context of $\Lambda$CDM cosmologies, it is natural to identify
the first dark matter condensations (``halos'') as the hosts of the
first sources of light.  A simplified picture for reionization is as
follows: the gas in dark matter halos cools and turns into ionizing
sources, which drive expanding ionized regions into the IGM.  The
volume filling factor of ionized regions then tracks the formation of
dark halos.  Eventually, the ionized regions percolate, and the
remaining neutral hydrogen in the IGM is quickly cleared away as the
ionizing background builds up.  From the point of view of the
large--angle CMB polarization anisotropies, what matters most is the
free electron fraction $x_e=n_e/n_{\rm tot}$.  Accordingly, we now
briefly describe the various effects that should determine the
evolution of $x_e(z)$ at high redshifts.  For in--depth discussion, we
refer the reader to extended reviews (Barkana \& Loeb 2001; Loeb \&
Barkana 2001), and to reviews focusing on the roles of ${\rm H_2}$
molecules in reionization (Abel \& Haiman 2000), on the effect of
reionization on CMB anisotropies (Haiman \& Knox 1999), and on
progress in the last two years in studying reionization (Haiman 2003).

In the discussions that follow, it will be useful to distinguish dark
matter halos in three different ranges of virial temperatures, as
follows:
\begin{center}
\begin{tabbing}
\hspace{2cm} \= $100\,{\rm K}$   $\lsim$ \= $T_{\rm vir}$ \=  $\lsim  10^4\, {\rm K}$ \hspace{1cm} \= (Type II)\\
             \> $10^4\,{\rm K}$  $\lsim$ \> $T_{\rm vir}$ \>  $\lsim 2\times 10^5\,{\rm K}$        \> (Type Ia)\\
             \>                          \> $T_{\rm vir}$ \>  $\gsim 2\times 10^5\,{\rm K}$        \> (Type Ib)
\end{tabbing}
\end{center}

We will hereafter refer to these three different types of halos as
Type II, Type Ia, and Type Ib halos.  The motivation in distinguishing
Type II and Type I halos is based on ${\rm H_2}$ molecular vs atomic H
cooling (\S~\ref{subsec:h2}), whereas types ``a'' and ``b'' reflect
the ability of halos to allow infall and cooling of photoionized gas
(\S~\ref{subsec:efficiencies}).  While the actual values of the
temperatures, and the sharpness of the transition from one category to
the next is uncertain, as we argue below, each type of halo likely
plays a different role in the reionization history.  In short, Type II
halos can host ionizing sources only in the neutral regions of the
IGM, and only if ${\rm H_2}$ molecules are present; Type Ia halos can
form new ionizing sources only in the neutral IGM regions, but
irrespective of the ${\rm H_2}$ abundance, and Type Ib halos can form
ionizing sources regardless of the ${\rm H_2}$ abundance, and whether
they are in the ionized or neutral phase of the IGM.  In the following
sections, we provide a summary of the various important physical
effects. Much of this summary relies on existing literature; our
primary difference from existing ideas is in the description of the
transition from metal--free to normal stellar population
(\S~\ref{subsec:efficiencies}).

\begin{table*}[b]\small
\caption{\label{table:collapse} Collapse redshifts and masses of 1, 2
and 3$\sigma$ dark matter halos with virial temperatures of $T_{\rm
vir}=100$ and $10^4$K, reflecting the minimum gas temperatures required for
cooling by ${\rm H_2}$ and neutral atomic H, respectively.}
\begin{center}
\begin{tabular}{lrrr}
$T_{\rm vir}$(K) & $\nu$ & $z_{\rm coll}$ & $M_{\rm halo}({\rm M_\odot})$ \\
\hline
100    & 1   &  7    & $2\times10^5$ \\
       & 2   &  16   & $5\times10^4$ \\
       & 3   &  26   & $3\times10^4$ \\
$10^4$ & 1   &  3.5  & $4\times10^8$ \\
       & 2   &  9    & $1\times10^8$ \\
       & 3   &  15   & $6\times10^7$ \\
\hline
\multicolumn{4}{l}{} \\
\end{tabular}\\[12pt]
\end{center}
\end{table*}

\subsection{${\rm H_2}$ Molecule Formation}
\label{subsec:h2}

In $\Lambda$CDM cosmologies, structure formation is bottom--up: the
earliest nonlinear dark matter halos form at low masses.  Gas
contracts together with the dark matter only in dark halos above the
cosmological Jeans mass, $M_{\rm J}\approx 10^4$. However, this gas
can only cool and contract to high densities in somewhat more massive
halos, with $M\gsim M_{\rm H2} \equiv 10^5\msun[(1+z)/11]^{-3/2}$
(i.e. Type II halos), and only provided that there is a sufficient
abundance of ${\rm H_2}$ molecules, with a relative number fraction at
least $n_{\rm H2}/n_{\rm H}\sim 10^{-3}$ (Haiman, Thoul \& Loeb 1996;
Tegmark et al. 1997).  Because the typical collapse redshift of halos
is a strong function of their size, the abundance of ${\rm H_2}$
molecules is potentially the most important parameter in determining
the onset of reionization.  For example, 2$\sigma$ halos with virial
temperatures of $100$K appear at $z=16$, while 2$\sigma$ halos with
virial temperatures of $10^4$K (Type Ia halos, in which gas cooling is
enabled by atomic hydrogen lines) appear only at
$z=9$. Table~\ref{table:collapse} summarizes the collapse redshifts
and masses of dark halos at these two virial temperatures.  As a
result, the presence or absence of ${\rm H_2}$ in Type II halos makes
a factor of $\sim$2 difference in the redshift for the onset of
structure formation, and thus potentially effects the reionization
redshift and the electron scattering opacity $\tau$ by a similar
factor. In the absence of any feedback processes, the gas collecting
in Type II halos is expected to be able to form the requisite amount
of ${\rm H_2}$ (Haiman, Thoul \& Loeb 1996; Tegmark et al 1997).
However, both internal and external feedback processes can alter the
typical ${\rm H_2}$ abundance (see discussion in
\S~\ref{subsec:feedback} below).

\subsection{The Nature of Ionizing Sources}
\label{subsec:sourcetype}

A significant uncertainty is the nature of the ionizing sources
turning on inside halos collapsing at the highest redshifts. Three
dimensional simulations using adaptive mesh refinement (AMR; Abel,
Bryan \& Norman 2000, 2002) and smooth particle hydrodynamics (SPH;
Bromm, Coppi \& Larson 1999, 2002) techniques have followed the
contraction of gas in Type II halos at high redshifts to high
densities. These works have shown convergence toward a
temperature/density regime of ${\rm T \sim 200~{\rm K}\,}$, ${\rm n
\sim 10^{4} \, {\rm cm}^{-3}}$, dictated by the critical density at
which the excited states of ${\rm H_2}$ reach equilibrium population
levels.  These results have suggested that the first Type II halos can
only form unusually massive stars, with masses of at least $\sim 200
\, {\rm M_\odot}$, and only from a small fraction ($\lsim 0.01$) of
the available gas.  Such massive stars are effective producers of
ionizing radiation, making reionization easier to achieve.  In
addition, it is expected that the first massive stars, forming out of
metal--free gas, have unusually hard spectra (Tumlinson \& Shull 2000;
Bromm, Kudritzki \& Loeb 2001; Schaerer 2002), which is important for
possibly ionizing helium, in addition to hydrogen.  An alternative
possibility is that a similar fraction of the gas in the first Type II
halos forms massive black holes (Haiman \& Loeb 1997; Bromm \& Loeb
2002).  These early black holes can then accrete gas, acting as
``miniquasars'', and producing a hard spectrum extending to the soft
X--rays. These soft X--rays could be important in catalyzing ${\rm
H_2}$ formation (see discussion in \S~\ref{subsec:feedback} below).

\subsection{Efficiency Parameters and the Transition from Metal Free to ``Normal'' Stars}
\label{subsec:efficiencies}

Another fundamental question of interest is the efficiency at which
the ionizing source(s) associated with each halo eject ionizing
photons into the IGM.  This can be parameterized by the product
$\epsilon_* \equiv N_\gamma f_* f_{\rm esc}$, where $f_* \equiv
M_*/(\Omega_{\rm b}M_{\rm halo}/\Omega_m)$ is the fraction of baryons
in the halo that turns into stars; $N_\gamma$ is the mean number of
ionizing photons produced by an atom cycled through stars, averaged
over the initial mass function (IMF) of the stars; and $f_{\rm esc}$
is the fraction of these ionizing photons that escapes into the IGM.
It is difficult to estimate these quantities at high--redshifts from
first principles, but the numbers and discussion below can serve as a
useful guide.

Although the majority of the baryonic mass in the local universe has
been turned into stars (Fukugita, Hogan \& Peebles 1998), the global
star formation efficiency at high redshifts was likely lower.  To
explain a universal carbon enrichment of the IGM to a level of
$10^{-2}-10^{-3}~{\rm Z_\odot}$, the required efficiency, averaged
over all halos at $z\gsim 4$, is $f_*=2-20\%$ (Haiman \& Loeb
1997). However, the numerical simulations mentioned in \S~2.2 above
suggest that the fraction of gas turned into massive stars in Type II
halos is $f_*\lsim 1\%$.

The escape fraction of ionizing radiation in local starburst galaxies
is of order $\sim 10\%$.  The higher characteristic densities at
higher redshifts could decrease this value (Dove et al. 2000; Wood \&
Loeb 2000), although there are empirical indications that the escape
fraction in $z\sim3$ galaxies may instead be higher (Steidel, Pettini
\& Adelberger 2001). Radiation can also more readily ionize the local
gas, and escape from the small Type II halos which have relatively low
total hydrogen column densities ($\lsim 10^{17}~{\rm cm^{-2}}$).

The ionizing photon yield per proton for a normal Salpeter IMF is
$N_\gamma\approx 4000$. However, if the IMF consists exclusively of
massive $M\gsim 200\,{\rm M_\odot}$ metal--free stars, then $N_\gamma$
can be up to a factor of $\sim 20$ higher (Bromm, Kudritzki \& Loeb
2001; Schaerer 2002).  The transition from metal--free to a ``normal''
stellar population is thought to occur at a critical metallicity of
$Z_{\rm cr}\sim 5\times 10^{-4}{\rm Z_\odot}$ (Bromm et al. 2001). It
is natural to associate this transition with that of the assembly of
halos with virial temperatures of $>10^4$K (Type Ia halos) . Type II
halos are fragile, and likely blow away their gas and ``shut
themselves off'' after a single episode of (metal--free)
star--formation. They are therefore unlikely to allow continued
formation of stars with metallicities above $Z_{\rm crit}$.
Subsequent star--formation will then occur only when the deeper
potential wells of Type Ia halos are assembled and cool their gas via
atomic hydrogen lines. The material that collects in these halos will
then have already gone through a Type II halo phase and contain traces
of metals.

As we argue in \S~\ref{subsec:feedback} below, there exists an
alternative, equally plausible scenario.  Most Type II halos may not
have formed ${\rm any}$ stars, due to global ${\rm H_2}$
photodissociation by an early cosmic soft--UV background.  In this
case, the first generation of metal--free stars must appear in Type Ia
halos. Halos above this threshold can eject most of their
self--produced metals into the IGM, but, in difference from Type II
halos, can retain most of their gas (MacLow \& Ferrara 1999), and can
have significant episodes of metal--free starformation.  These halos
will also start the process of reionization by driving expanding
ionization fronts into the IGM. The metals that are ejected from Type
Ia halos will reside in these photoionized regions of the IGM. As
discussed in \S~{ref:feedback} below, photoionization heating in these
regions suppresses gas infall and cooling, causing a pause in the
formation of new structures, until larger dark matter halos, with
virial temperatures of $T_{\rm vir}\gsim 2\times 10^5$K (Type Ib
halos) are assembled.  The material that collects in Type Ib halos
will then have already gone through a previous phase of
metal--enrichment by Type Ia halos, and it is unlikely that Type Ib
halos can form significant numbers of metal--free stars.

The effect of a metal--free to a normal stellar population on the
reionization history was studied recently by Wyithe \& Loeb (2003) and
Cen (2003).  Both of these works assumed that rather than being tied
either to a Type II$\rightarrow$Ia or Type Ia$\rightarrow$Ib halo
transition, the switch from metal--free to normal stars occurs
abruptly at a critical redshift $z_{\rm cr}$. This would be justified
in the picture of a uniformly increasing metal--enrichment of the IGM,
and would lead to qualitatively different reionization histories from
the ones we obtain below.

\subsection{Density Fluctuations (Clumpiness) of the IGM}
\label{subsec:clumping}

As the ionizing fronts propagate into the IGM, their expansion rate at
high redshifts is limited by the mean recombination rate in the
ionized gas (rather than simply by the ionizing luminosity). The
recombination rate per unit time and volume can be written as
$\alpha_{\rm B}C_{\rm HII} \langle n_{\rm HII} \rangle^2$, where
$\alpha_{\rm B}$ is the recombination coefficient of neutral hydrogen
to its excited states ($= 2.6\times10^{-13}~{\rm cm^3~s^{-1}}$ at
$T=10^4$K), $\langle n_{\rm HII} \rangle$ is the mean number density
of ionized hydrogen, and $C_{\rm HII}\equiv \langle n_{\rm HII}^2
\rangle / \langle n_{\rm HII} \rangle^2$ is the mean clumping factor
of ionized gas.  The clumping factor can be estimated by numerical
simulations (Gnedin \& Ostriker 1997), and in semi--analytic models
(Madau et al. 1999; Chiu \& Ostriker 2000; Miralda-Escud\'e et
al. 2000; Haiman, Abel \& Madau 2001; Benson et al. 2001; Wyithe \&
Loeb 2003).  It is expected that the clumping factor rises from a
value near unity to a few tens, as structures go increasingly
nonlinear, between redshift redshifts $z=10-20$.  Numerical
simulations predict $C_{\rm HII}=1-10$ at $z>10$, but at high
redshifts, where numerical simulations cannot resolve the Jeans mass,
and the effective clumping can be increased by fluctuations on small
scales (Haiman, Abel \& Madau 2000), to $C_{\rm HII}\approx 40$ at
$z\sim 10$.  Finally, we note that whenever recombinations are fast
enough to set the expansion rate of the ionization fronts, which is
indeed the case in the models we consider below, it is essentially the
linear combination $\epsilon_*/C_{\rm HII}$ that determines the
evolution of the total ionized volume fraction.

\subsection{Feedback Effects}
\label{subsec:feedback}

Several feedback effects are likely to be important for reionization.
There can be significant {\it internal} feedback in or near each
ionizing source, due to the presence of supernovae (Ferrara 1998), or
of the radiation field (Omukai \& Nishi 1999; Ricotti, Gnedin \& Shull
2002), on the local ${\rm H_2}$ chemistry. The net sign of these
effects is difficult compute, as it depends on the source properties
and spectra, and on the detailed density distribution internal and
near to the sources. For practical purposes of computing the global
reionization history, we may, however, think of any internal feedback
effect as regulating the efficiency parameter $\epsilon_*$ defined
above.

Since the universe is optically thin at soft UV (<13.6eV), and soft
X-ray ($\gsim$1keV) photon energies, radiation from the earliest Type
II halos can build up global backgrounds in these bands, and provide
prompt {\it external} feedback on the formation of subsequent
structures, which are easier to follow.  In particular, soft UV
photons can photodissociate ${\rm H_2}$ molecules, while X-rays can
promote ${\rm H_2}$ formation (Haiman, Abel \& Rees 2000).  The net
sign of this feedback depends on the spectrum (with a critical
X-ray/UV energy flux ratio of a few percent) of the earliest sources
(Haiman, Abel \& Rees 2000; Ciardi, Ferrara \& Abel 2000; Machacek,
Bryan \& Abel 2001, 2003; Glover \& Brandt 2003).

A second type of important feedback is that photo--ionized regions are
photo-heated to a temperature of $\gsim 10^4$K, with a corresponding
increase in the Jeans mass in these regions. As a result, gas from
existing Type II halos is photo--evaporated (Shapiro et al. 1998;
Barkana \& Loeb 1999), and if the gas is kept photoionized, then Type
Ia halos are subsequently prohibited from collecting gas at their
centers.  Gas infall and cooling in the ionized zones is suppressed in
halos with virial temperatures up to $\sim 2\times 10^5$K (Efstathiou
1992; Thoul \& Weinberg 1995; Navarro \& Steinmetz 1997).  Type Ia
dark matter halos that form in already ionized regions are therefore
effectively excluded from contributing to reionization; in the ionized
regions, new ionizing sources will have to wait until Type Ib halos
build up further down in the hierarchy of structure formation.

\section{Models of Reionization}
\label{sec:models}

In this section, we describe our semi--analytical model of the
reionization process, incorporating the above effects in the simplest
possible way.  In these models, we track the total volume of ionized
regions, assuming that ionizing sources are located inside virialized
dark matter halos. Each source creates an ionized Str\"omgren region,
which expands into the IGM at a rate dictated by the source
luminosity, and by the background IGM density and clumping factor.

We adopt the mass function of virialized dark matter halos from the
N--body simulations of Jenkins et al. (2001).  The comoving number
density $(dn/dM)dM$ of halos at redshift $z$ with mass $M\pm dM/2$ is
given by 
\beq \frac{dn}{dM}(z,M)= 0.315 \frac{\rho_0}{M}
\frac{1}{\sigma_M} \frac{d\sigma_M}{dM}
\exp\left[-\left|0.61-\log(D_z\sigma_M)\right|^{3.8}\right],
\label{eq:dndm}
\eeq where $\sigma_M$ is the r.m.s. density fluctuation, computed on
mass--scale $M$ from the present--day linear power spectrum
(Eisenstein \& Hu 1999), $D_z$ is the linear growth function, and
$\rho_0$ is the present--day total mass density.

The gas inside each virialized object is shock--heated to the virial
temperature,
\beq
T_{\rm vir}\approx 1800
\left(\frac{M}{10^6{\rm M_\odot}}\right)^{2/3}
\left(\frac{1+z}{21}\right)
\left(\frac{\Omega_0}{0.3}\right)^{1/3}
\left(\frac{h}{0.7}\right)^{2/3}
\left(\frac{\mu}{1.22}\right)
\,{\rm K}.  
\label{eq:tvir}
\eeq where $\mu$ is the mean molecular weight ($\mu=0.60$ for ionized
gas collecting in Type Ia,b halos, and $\mu=1.22$ for neutral gas of
primordial composition, relevant for gas collecting in the shallow
potentials of Type II halos).

\begin{figure}[t]
\plotone{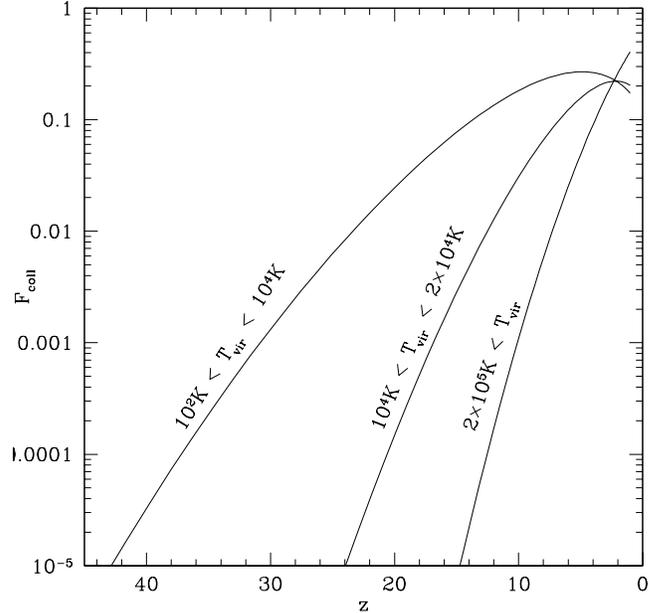}
\caption{The evolution of the collapsed fraction of all baryons in
halos in three different virial temperature ranges.  The three
different virial temperature ranges represent halos that can form
ionizing sources (1) only in the neutral regions of the IGM, and only
if ${\rm H_2}$ molecules are present, (2) only in the neutral IGM
regions, irrespective of the ${\rm H_2}$ abundance, and (3),
irrespective of the ${\rm H_2}$ abundance, and both in the ionized or
neutral phases of the IGM.}
\label{fig:fcol} 
\end{figure}

In order to address the two global radiative feedback effects
described in \S~2.5 above, we separately define, and track, the total
fraction of the mass in the universe that is condensed into Type II,
Ia, and Ib halos:
\beq F_{\rm coll,II}(z)\equiv \frac{1}{\rho_0} \int_{M_{\rm
II}}^{M_{\rm Ia}} dM \left(\frac{dn}{dM}\right) M,
\label{eq:fh2}
\eeq
\beq
F_{\rm coll,Ia}(z)\equiv 
\frac{1}{\rho_0}
\int_{M_{\rm Ia}}^{M_{\rm Ib}} dM \left(\frac{dn}{dM}\right) M,
\label{eq:fhi}
\eeq 
and
\beq
F_{\rm coll,Ib}(z)\equiv 
\frac{1}{\rho_0}
\int_{M_{\rm Ib}}^\infty dM \left(\frac{dn}{dM}\right) M,
\label{eq:fhii}
\eeq where the halo masses $M_{\rm II}=M_{\rm II}(T_{\rm II},z)$,
$M_{\rm Ia}=M_{\rm Ia}(T_{\rm Ia},z)$, and $M_{\rm Ib}=M_{\rm
Ib}(T_{\rm Ib},z)$ are obtained from equation~(\ref{eq:tvir}) for the
temperatures $T_{\rm II}= 100$\,K, $T_{\rm Ia}= 10^4$\,K and $T_{\rm
Ib}= 2\times 10^5$\,K.  Our results below are insensitive to the upper
limit in equation~(\ref{eq:fhii}).  The collapsed fractions are shown
in Figure~\ref{fig:fcol}. The time derivatives of
equations~(\ref{eq:fh2})-(\ref{eq:fhii}), in turn, represent the rates
at which mass is condensing into nonlinear halos. At low redshifts,
the derivatives can turn negative; which represents the decrease in
the number of halos as they merge into larger systems. This does not
significantly effect our results at the high redshifts we consider
$z\gsim 6$.

We assume that each halo represents a single cosmological ``reionizing
source'', which drives an expanding ionized region into the IGM.  For
each initial source redshift, we solve the equation of motion for the
ionization front, $R_i$,
\begin{equation}
\frac{dR_i^3}{dt}= 3H(z) R_i^3 +
 \frac{3 \dot N_\gamma}{4\pi \langle n_H \rangle}
 - C_{\rm HII} \langle n_H \rangle \alpha_B R_i^3,
\label{eq:Ri}
\end{equation}
where $H(z)$ is the Hubble constant at $z$, $\alpha_{\rm B}=
2.6\times10^{-13}~{\rm cm^3~s^{-1}}$ is the recombination coefficient
of neutral hydrogen to its excited states at $T=10^4$K, and $C_{\rm
  HII}$ is the mean clumping factor of ionized gas within $R_i$ as
discussed in \S~2.4 above. The first term in equation~(\ref{eq:Ri})
accounts for the Hubble expansion, the second term accounts for the
ionizations by newly produced photons, and the last term describes
recombinations (Shapiro \& Giroux 1987; Haiman \& Loeb 1997).

In equation~(\ref{eq:Ri}), $\dot N_\gamma$ represents the emission
rate of ionizing photons from a single halo.  For a stellar population
with a Salpeter IMF (initial mass function) undergoing a burst of star
formation at a metallicity equal to $2\%$ of the solar value
(Leitherer et al. 1999), we find that the rate is well approximated by
\beq \dot N_\gamma(t) = \left\{\matrix{ \dot N_0 \hfill&(t \leq
10^{6.5}~{\rm yr})\hfill\cr \dot N_0 (t/10^{6.5}{\rm yr})^{-4.5}
\hfill&(t > 10^{6.5}~{\rm yr})\hfill,\cr}\right.
\label{eq:rate}
\eeq where $\dot N_0=3.7\times 10^{46}~{\rm s^{-1}~M_\odot^{-1}}$ (per
${\rm M_\odot}$ of stellar mass).  Over the lifetime of the
population, this produces $\approx 4000$ ionizing photons per stellar
proton.  In practice, the injection of ionizing photons is nearly
instantaneous compared to the expansion timescale even at redshift
$z\sim 20$, and our results are insensitive to the adopted lightcurve
(an assumption that remains justified for an IMF biased to more
massive stars).

In order to investigate the various physical effects outlined in \S~2,
we define separate star--formation efficiencies, $\epsilon_{\rm II}$,
$\epsilon_{\rm Ia}$, and $\epsilon_{\rm Ib}$ for Type II, I, and Ib
halos, respectively.  In order to allow for a negative radiative
feedback by the soft UV background on the ${\rm H_2}$ abundance, we
consider a redshift $z_{\rm uv}$ below which $\epsilon_{\rm II}=0$, so
that Type II halos are excluded from contributing to reionization.  In
reality, rather than dropping to zero, the efficiency in Type II halos
should roughly remain at a constant ``steady state'' value, set by the
level of starformation that can just be maintained against the ${\rm
H_2}$ photo--dissociation implied by this starformation (Haiman, Abel
\& Rees 2000).  Retaining this level of starformation in our models
would have a relatively small effect on our results.

For the sake of simplicity, we adopt a constant clumping factor
$C_{\rm HII}$.  While the clumping factor does evolve with redshift,
at $z\gsim 10$, this could be absorbed as a factor of $\sim 10$
decrease in the efficiencies $\epsilon$ down to this redshift.
Including the evolution of the clumping factor with redshift would be
necessary in order to model the evolution of the {\it neutral}
fraction at $5.5\lsim z\lsim 6.5$, necessary to interpret its sharp
rise in the observed high--redshift quasar spectra.  However,
semi--analytical models (Haiman \& Loeb 1997, 1998) and numerical
simulations (Gnedin \& Ostriker 1997; Gnedin 2001; Razoumov et
al. 2002; ) have shown that even though the entire process of
reionization can last for an extended period, the rapid evolution in
the neutral fraction, caused by the sudden buildup of the radiation
background once the ionized regions percolate, follows the percolation
epoch within a short redshift interval $\Delta z<1$. We will require
only that our model indicate a ``second reionization'' epoch at $z\sim
7$, and defer detailed modeling of the neutral fraction to a future
paper.

Combining the abundance of the ionizing sources (eq.~\ref{eq:dndm})
with the ionized volume associated with each source (eq.~\ref{eq:Ri}),
we can compute the volume filling fraction of the ionized regions as a
function of redshift.  The ionized volume associated with a halo of
mass M scales linearly with its mass and with the efficiency factor
$\epsilon$), namely $V_{\rm HII}(z_{\rm on},z,M)=(4\pi R_{\rm i}^3/3)
= \epsilon M\times {\tilde V}_{\rm HII}(z_{\rm on},z)$, where ${\tilde
V}_{\rm HII}(z_{\rm on},z)$ is the ionized volume per unit mass and
unit efficiency, and $R_{\rm i}$ is the solution of
equation~(\ref{eq:Ri}).  The HII filling factor $F_{\rm HII}(z)$ is
then obtained from the collapsed gas fractions as
\begin{eqnarray}
\nonumber
F_{\rm HII}&&(z)=\rho_{\rm b}(z)
\int_{\infty}^{z}dz^{\prime} 
\left\{
  \epsilon_{\rm Ib} \frac{dF_{\rm coll,Ib}}{dz} (z^{\prime}) 
+ \left[1-F_{\rm HII}(z^\prime)\right]\right.\times\\
&&\left.\left[  \epsilon_{\rm Ia} \frac{dF_{\rm coll,Ia}}{dz} (z^{\prime}) 
        + \epsilon_{\rm II}  \frac{dF_{\rm coll,II}}{dz} (z^{\prime})
 \right]
\right\}
{\tilde V}_{\rm HII}(z^{\prime},z),
\label{eq:filling}
\end{eqnarray}
where $\rho_{\rm b}=\Omega_{\rm b}\rho_{\rm crit}$ is the average
baryonic density and $\rho_{\rm crit}$ is the critical density of the
universe.  In the second term on the right hand side of
equation~(\ref{eq:filling}), we have explicitly included a factor
$(1-F_{\rm HII})$, which takes into account the fact that new ionizing
sources should appear in Type II and Type Ia halos only in regions
which have not yet been ionized.

Equations~(\ref{eq:dndm}-\ref{eq:filling}) determine the evolution of
the ionized fraction. Note that equation~(\ref{eq:filling}) is
implicit, and, in practice, has to be solved numerically by starting
at high redshift ($z=60$), and taking sufficiently small negative
redshift steps (we found $\Delta z\approx 0.001$ to be adequate in
most of the models we studied).  In summary, our model has five
parameters: the clumping factor $C_{\rm HII}$, the overall
efficiencies, $\epsilon_{\rm II}$, $\epsilon_{\rm Ia}$, $\epsilon_{\rm
Ib}$, and the redshift $z_{\rm uv}$ at which ${\rm H_2}$ dissociative
feedback sets in.  Since variations in the clumping factor are nearly
degenerate with changes in the efficiency parameters, in all of our
models we fix $C_{\rm HII}=10$ (alternatively, adopting a
redshift--dependent clumping factor from the works discussed in
\S~\ref{subsec:clumping} would only case a ``tilt'' in the
reionization histories we derive below).  We therefore have four
``free'' parameters.  Current CMB constraints are not sufficiently
accurate to converge on a ``best--fit'' model, and our main purpose
here is to be able to study the various physical effects outlined in
\S~\ref{sec:physics} above.  Nevertheless, the hope (as we attempt to
demonstrate in \S~\ref{sec:cmb} below) is that eventually, CMB
polarization data {\it will} provide constraints on parameters of
models similar to the one described here, while simultaneously
delivering constraints on cosmological parameters.

\begin{figure}[t]
\plotone{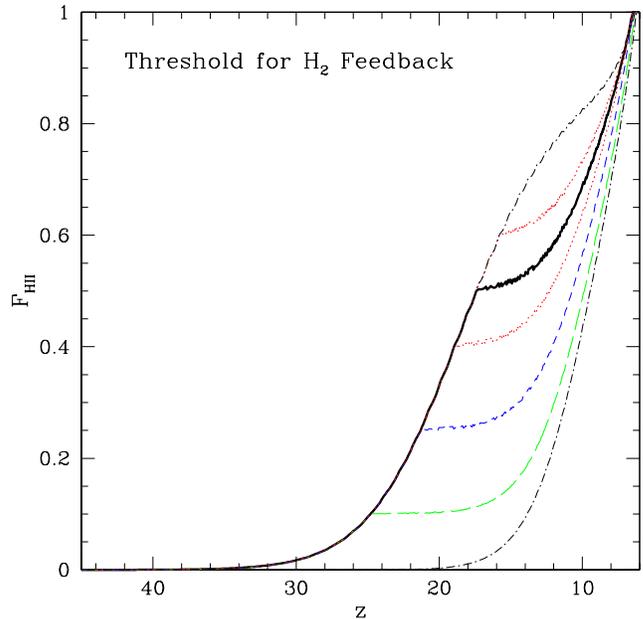}
\caption{The evolution of the ionized fraction of all baryons.  The
thick solid curve corresponds to our fiducial model described in
\S~\ref{subsec:fiducial}. The other curves indicate variations in the
importance of ${\rm H_2}$ photodissociation feedback, from no feedback
(top dot--dashed curve), to strong feedback that effectively excludes
Type II halos from contributing to reionization (bottom dot--dashed
curve).  The pair of dotted curves bracket a range $\Delta\tau=\pm0.01$
around the fiducial model; the short and long--dashed curves
correspond to reductions in $\tau$ by $\Delta\tau=0.03$ and $0.06$,
respectively (see \S~\ref{subsec:zuv} for discussion).}
\label{fig:xe1} 
\end{figure}

\section{Possible Reionization Histories}
\label{sec:histories}

In this section, we present a suite of models, designed to investigate
the various effects summarized in \S~2 above. Note that more realistic
calculations would explicitly have to couple the star formation
efficiency with radiative feedback processes (as attempted in part by
Haiman, Abel \& Rees 2000, Glover \& Brandt 2003, and Cen 2003).  In
the present models, we explicitly track only the ionized fraction of
hydrogen.  For the purpose of computing the electron scattering
optical depth, we assume that the HII and HeII fractions are
identical, so that the extra electrons from HeII increase the opacity
by a factor of 1.08. A HeII to HeIII transition appears to be taking
place at the low redshift $z\sim 3$ (Miralda-Escud\'e 2001). While
some HeIII could be present at high redshifts, for realistic spectral
slopes of the ionizing sources, the extra electrons from HeIII regions
add corrections to $\tau$ at the $\lsim 10\%$ level.

\begin{figure}[t]
\plotone{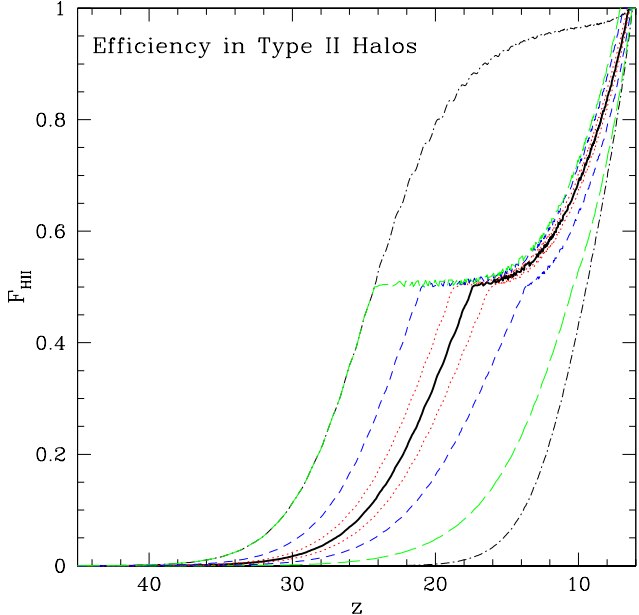}
\caption{The evolution of the ionized fraction of all baryons, with
the thick solid curve showing the fiducial model as in
Figure~\ref{fig:xe1}.  The other curves indicate variations in the
efficiency parameter $\epsilon_{\rm II}$ in Type II halos. As in
Figure~\ref{fig:xe1}, the pairs of dotted, short--dashed, and
long--dashed curves bracket a range of $\Delta\tau=\pm0.01$, $0.03$,
and $0.06$ around the fiducial model.  The required changes in the
efficiencies are described in the text.  The bottom dot--dashed curve
assumes Type II halos do not form stars, and the top dot--dashed curve
assumes that there is no ${\rm H_2}$ photodissociation feedback (see
\S~\ref{subsec:epsII} for discussion).}
\label{fig:xe2} 
\end{figure}

\subsection{Fiducial Model}
\label{subsec:fiducial}
As discussed above, in order to match inferences from the $z<6$ quasar
spectra, we require that a percolation takes place at redshift $z\sim
7$.  A second motivation for requiring a percolation near $z\sim 7$ is
the high observed temperature of the IGM at $z\sim 4$: if the
percolation occurred too early, the IGM would cool to temperatures that
may be too low by $z\approx4$ (Hui \& Haiman 2003).  We find that
this effectively fixes the efficiencies $\epsilon_{\rm
Ia}=\epsilon_{\rm Ib}=80$ (if they are postulated to have the same
value).  Accordingly, in our fiducial model, we adopt these
values. Note that these values are quite reasonable: for a normal
stellar population, $N_\gamma=4000$, and we may break down the
efficiency into $f_*=0.2$ and $f_{\rm esc}=0.1$.  With $\epsilon_{\rm
Ia}$ and $\epsilon_{\rm Ib}=80$ fixed, we have only $z_{\rm uv}$ and
$\epsilon_{\rm II}$ to vary. In our fiducial model, Type II halos host
metal--free stars from a small fraction ($f_*=0.005$) of the available
gas, but the stars are, on average, 10 times more efficient ionizing
photon producers ($N_\gamma=40000$), and all of their ionizing
radiation escapes into the IGM ($f_{\rm esc}=1$).  This results in the
overall efficiency of $\epsilon_{\rm II}=200$.  We then finally adopt
a redshift $z_{\rm uv}=17$ for the onset of ${\rm
H_2}$--photodestruction, which allows Type II halos to ionize $\sim$
half of the volume of the IGM.  By construction, this model produces
an optical depth $\tau=0.17$, the central value in the range allowed
by {\it WMAP}. In what follows, we separately vary each parameter --
this will explicitly clarify the dependence of the reionization
history and of $\tau$ on each of our parameters.

\subsection{When does ${\rm H_2}$ dissociation feedback set in?}
\label{subsec:zuv}

The redshift at which any ${\rm H_2}$ dissociation sets in is
determined by the spectra of the light sources in Type II halos, and
the intensity of the X--ray and soft UV backgrounds. According to
Haiman, Abel \& Rees (2000), the choice of $z_{\rm uv}=17$, resulting
in Type II halos ionizing $\sim$half of the IGM volume, would be
consistent with an X--ray/UV background energy flux ratio of a few
percent.  We here simply parameterize this feedback with the redshift
of its onset, $z_{\rm uv}$. The reionization history in our fiducial
model is shown by the thick solid curve in Figure~\ref{fig:xe1}.  The
other curves in this figure show variations in $z_{\rm uv}$ away from
the fiducial model, which correspond to changes $\Delta\tau=0.01$,
$0.03$, and $0.06$.  The level of the constraint that will be provided
by Planck corresponds approximately to $\Delta\tau=0.01$ (although
Planck will have information beyond a measurement of $\tau$, see
discussion in \S~\ref{sec:cmb} below and in paper II).  As the dotted
curves in the figure show, at the level of the Planck constraint,
$z_{\rm uv}$ could be determined to a precision of $\Delta z_{\rm uv}
\approx \pm 1.5$.  Note that even in the absence of any ${\rm H_2}$
destruction (upper dot--dashed curve), the opacity can only be
increased to $\tau \sim 0.185$.

\begin{figure}[t]
\plotone{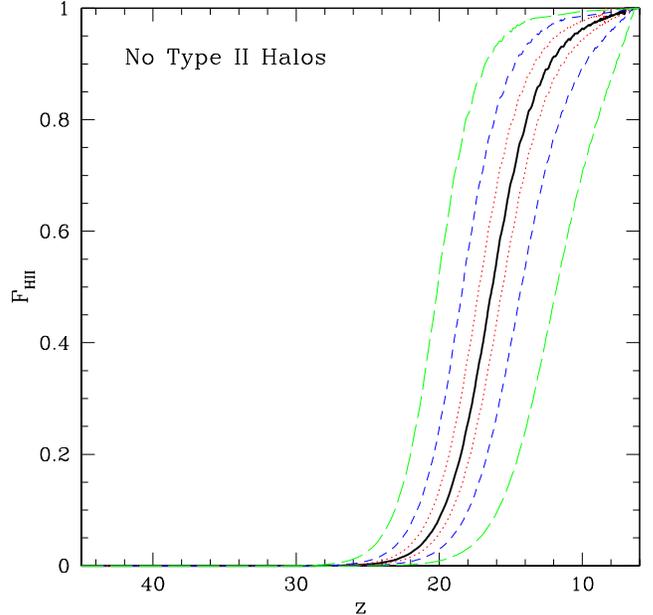}
\caption{This figure assumes that ${\rm H_2}$ feedback effectively
shuts off starformation in all Type II halos, and therefore Type Ia
halos form metal free stars, but Type Ib halos form normal stars.  The
thick solid curve shows a new fiducial model that has the
$\tau=0.17$. This requires increasing the efficiencies in Type Ia
halos by a factor of 60 relative to the fiducial models in
Figures~\ref{fig:xe1} and \ref{fig:xe2}.  The pairs of dotted,
short--dashed, and long--dashed curves bracket a range of
$\Delta\tau=\pm0.01$, $0.03$, and $0.06$ around this new fiducial
model. Note we keep the efficiency in Type II halos fixed, and
percolation in all models is achieved only at $z\sim 6$. This can lead
to consistency with inferences from $z\sim 6$ SDSS quasar spectra
(see \S~\ref{subsec:notypeII} for discussion).}
\label{fig:xe4feed} 
\end{figure}

\subsection{How different is the overall efficiency in Type II halos?}
\label{subsec:epsII}
In our fiducial model, we have adopted a ratio $\epsilon_{\rm
II}/\epsilon_{\rm Ia}=2.5$, determined by the transition from
metal--free to normal stars. In Figure~\ref{fig:xe2}, we consider
variations in $\epsilon_{\rm II}$.  We show results for values of
$\epsilon_{\rm II}$ that result in changes of $\pm\Delta\tau=0.01$,
$0.03$, and $0.06$ (dotted, short--dashed, and long--dashed curves,
respectively).  In each case, we adjust the value of $z_{\rm uv}$ so
that the Type II halos always ionize $\sim$ half of the volume (this
approximately mimics fixing the radiation intensity at which ${\rm
H_2}$ feedback sets in).  The value of $\epsilon_{\rm II}$ for these
six curves, from bottom to top, are $\epsilon_{\rm II}=20, 80, 120,
280, 520$, and $1400$.  The last value still represents a relatively
modest factor of $\sim 7$ increase in the combined efficiencies, and
is within the range of uncertainties.  The Planck--level constraints
represent $\sim 50\%$ changes in $\epsilon_{\rm II}$. The bottom
dot--dashed curve shows the case of no starformation in Type II halos
($\epsilon_{\rm II}=0$; this is equivalent to $z_{\rm uv}\gsim 35$,
shown by the bottom curve in Figure~\ref{fig:xe1}).  The top
dot--dashed curve shows the highest efficiency case, but assuming no
${\rm H_2}$ dissociation. In this model, $\tau=0.29$, which would be
ruled out at the $\sim 3\sigma$ level by the current {\it WMAP}
results.

\subsection{Are ${\rm H_2}$ molecules required to produce large $\tau$ ?}
\label{subsec:notypeII}
In Figure~\ref{fig:xe4feed}, we assume that ${\rm H_2}$ feedback
effectively shuts off starformation in all Type II halos. As a result,
Type Ia halos form metal--free stars, and, motivated by the arguments
in \S~\ref{subsec:efficiencies}, we assume that the transition to
normal stellar populations took place in Type Ib halos (and we set
$\epsilon_{\rm Ib}=80$).  We then revise our fiducial model and find
that $\epsilon_{\rm Ia}=4800$ is required to produce $\tau=0.17$,
implying an overall efficiency difference of a factor of 60 between
Type Ia and Ib halos (shown by the solid curve).  The main conclusion
from Figure~\ref{fig:xe4feed} is that $\tau=0.17$ can be achieved
without ${\rm H_2}$ molecules, but this requires a boost in
efficiencies by this factor. This would require, in turn, in addition
to the maximum factor of $\sim 20$ caused by switching from
metal--free to normal stars, that the escape fraction or
star--formation efficiency in Type Ia halos was a factor of $\sim 3$
higher (or else that the clumping factor decreases by the same factor
from $z\sim 15$ to $z\sim 10$).  Another important feature apparent
from Figure~\ref{fig:xe4feed} is that the reionization history is
qualitatively different from those shown in Figures~\ref{fig:xe1} and
\ref{fig:xe2}.  As we discuss in \S~\ref{sec:cmb} below, future CMB
polarization measurements can distinguish between the two scenarios at
high significance.  As in the previous figures, we also show the
$\pm\Delta\tau=0.01$, $0.03$, and $0.06$ ``contours'', which
correspond (bottom to top curves) to $\epsilon_{\rm Ia}=$ 480, 1600,
3200, 8000, 16000, 48000.  Planck--level constraints on $\epsilon_{\rm
Ia}=$ are at the $\sim 50\%$ level.  It is worth pointing out that in
the models shown in Figure~\ref{fig:xe4feed}, Type Ia sources are
self--limiting by photo--heating feedback (shown by the flattening of
the ionization histories below $z\lsim 12$) and it is the Type Ib
sources that eventually cause percolation at $z\sim 6.5$ in all
models, regardless of the efficiency in Type Ia sources.

\begin{figure}[t]
\plotone{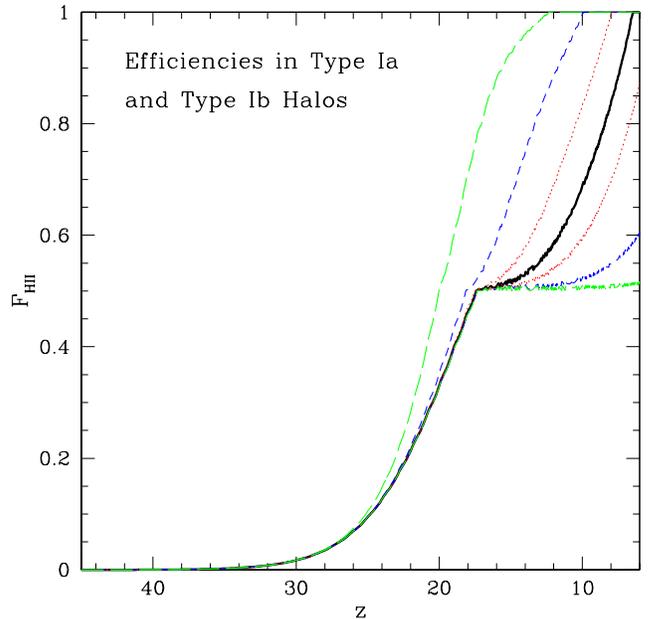}
\caption{The evolution of the ionized fraction of all baryons, with
the thick solid curve showing the fiducial model as in
Figure~\ref{fig:xe1}.  The other curves indicate variations in the
star formation efficiency in Type Ia and Ib halos .  The pairs of
dotted, short--dashed, and long--dashed curves bracket a range of
$\Delta\tau\pm0.01$, $0.03$, and $0.06$ around the fiducial model, as
in Figure~\ref{fig:xe2} (see \S~\ref{subsec:percolate} for
discussion). }
\label{fig:xe3} 
\end{figure}

\subsection{Constraints on efficiencies from the percolation epoch}
\label{subsec:percolate}
In our models, Type Ib halos form last, and must be responsible for
the eventual percolation of the ionized regions. As argued above, to
be consistent with inferences from $z\lsim 6$ quasar spectra, this
percolation epoch cannot be at $z\gsim 7$.  It is interesting to ask,
then: what is the allowed efficiency range in Type Ib halos?  To
address this question, we consider the fiducial model defined in
\S~\ref{subsec:fiducial}, and we vary $\epsilon_{\rm Ia}=\epsilon_{\rm
Ib}$. The resulting reionization histories for variations that cause
$\pm\Delta\tau=0.01$, $0.03$, and $0.06$ are shown in
Figure~\ref{fig:xe3}.  The figure shows that in high--efficiency
models (corresponding to $\Delta\tau\gsim 0.03$, or
$\epsilon_{Ib}\gsim 1200$), percolation occurs at too high redshifts.
Similarly, models with too low efficiencies ($\Delta\tau\lsim -0.01$,
or $\epsilon_{Ib} \lsim 30$) are ruled out as they fail to reionize
the universe by redshift $z\sim 6$.  The low--redshift observations by
themselves therefore constrain the efficiencies in Type Ib halos to
within a range of a factor of $\sim 30$.

\begin{figure}[t]
\plotone{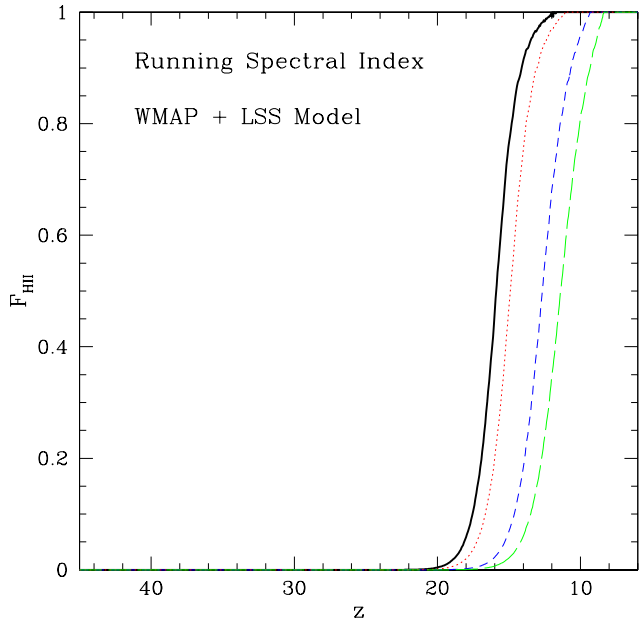}
\caption{This figure replaces the best fit power law {\it WMAP} cosmological
model by the best fit model that allows a running spectral index and
combines {\it WMAP} and other datasets (Table 10 in Spergel et al. 2003).
In this model, small--scale power is significantly reduced.  The thick
solid curve shows a new fiducial model that assumes no star--formation
in Type II halos, and has $\tau=0.17$. This requires boosting the
efficiency by an unrealistic factor of $3000$ relative to our
fiducial model in the power--law cosmology.  The dotted,
short--dashed, and long--dashed curves correspond to lower
efficiencies, and reduced values of $\tau$ by $\Delta\tau=-0.01$,
$-0.03$, and $-0.06$, respectively (see \S~\ref{subsec:running} for
discussion).}
\label{fig:xe4run} 
\end{figure}

\subsection{The running spectral index {\it WMAP} model}
\label{subsec:running}
When combined with other large scale structure data, the {\it WMAP}
results favor, at the $\sim 2\sigma$ level, a non--scale--invariant
primordial power spectrum with a running spectral index (Spergel et
al. 2003).  This model has much less small scale power than the
best--fit {\it WMAP}--only model. It is interesting to ask whether
$\tau=0.17$ can still be achieved in the running index model without
${\rm H_2}$ molecular cooling (Spergel et al. 2003).  To address this
question, we consider the fiducial model defined in
\S~\ref{subsec:notypeII}, but use the cosmological parameters listed
in Table 10 of Spergel et al. (2003).  We then increase $\epsilon_{\rm
Ia}$ until $\tau=0.17$ is achieved. In Figure~\ref{fig:xe4run}, we
show the reionization history in this model, which requires boosting
the efficiencies by a factor of 3000, to $\epsilon_{\rm
Ia}=2.5\times10^5$.  This model must unrealistically assume that all
the gas turns into stars (and gets recycled $\sim$3 times), all of the
ionizing radiation escapes into the IGM, the clumping factor is
$C_{\rm HII}=1$, and adopt the maximum factor of $\sim 20$ in the
increase in ionizing photon production rates in massive, metal--free
stars.  The dotted, short--dashed, and long--dashed curves assume
$\epsilon_{\rm Ia}=80000, 20000$, and 9000, corresponding to
decreasing $\tau$ by $\Delta\tau=-0.01, -0.03$, and -0.06,
respectively, and showing that the lack of molecules could still be
consistent with the lower range of $\tau$ allowed in the running index
model.  Allowing for Type II halos (assuming no ${\rm H_2}$ feedback)
we find that a relatively more realistic $\epsilon_{II}=4000$ is
required to produce $\tau=0.17$.

\subsection{Summary}
\label{subsec:historysummary}

The physics of reionization is rich in features that can naturally
lead to distinctive ionization histories.  These features can arise
because of (1) the different types of coolants in Type II and Type I
halos, (2) the different response of different halos to radiative
feedback on the ${\rm H_2}$ chemistry, and to photoionization feedback
on gas infall, and (3) the different efficiencies expected in
metal--free and normal stellar populations.  We generically find that
no simple model fits both the {\it WMAP} data and the constraint from
$z\lsim 6$ quasar spectra requiring that an episode of percolation
occur near $z\lsim7$.  Nevertheless, a range of interesting, physically
motivated reionization histories is allowed by current data. We find
that the ionized fraction $F_{\rm HII}(z)$ in these models evolves
monotonically with redshift, except under the assumption of a
uniformly increasing metal--enrichment of the IGM, and a sudden
transition from metal--free to ``normal'' stars at some redshift
$z_{\rm cr}$ (see discussion in next section, and also Wyithe \& Loeb
2003 and Cen 2003).

\begin{figure}[t]
\plotone{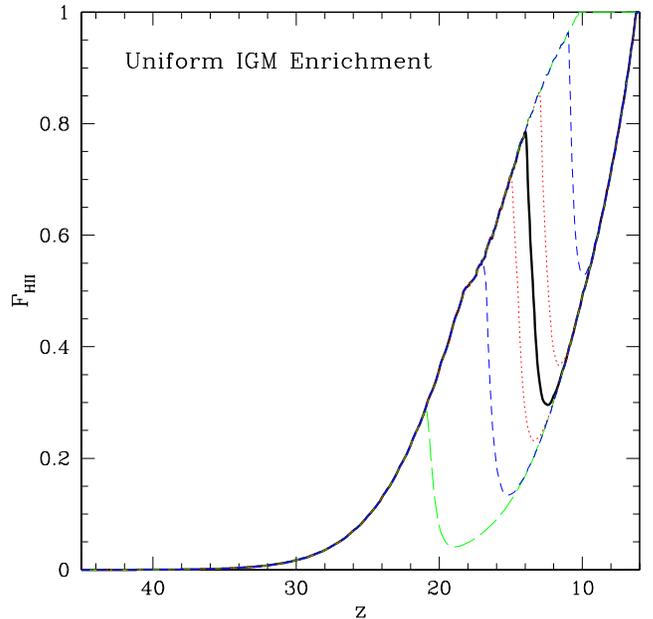}
\caption{This figure addresses the hypothesis that a transition from
metal free to normal star formation takes place abruptly at a fixed
redshift $z_{\rm cr}$, accompanied by a sudden decrease by a factor of
$f_{\rm cr}$ in the ionizing photon production rates.  The thick solid
curve shows a model that has $\tau=0.17$; this is achieved by $z_{\rm
cr}=14$ and $f_{\rm cr}=20$. The pairs of dotted, short--dashed, and
long--dashed curves bracket a range of $\Delta\tau\pm0.01$, $0.03$,
and $0.06$ around this model by changing $z_{\rm cr}$.}
\label{fig:xe5} 
\end{figure}

\section{Constraints From Future CMB Anisotropy Measurements}
\label{sec:cmb}

All models shown in Figures~\ref{fig:xe1}-\ref{fig:xe4run} in the
previous section are consistent with the optical depth
$\tau=0.17\pm0.04$ measured in the current {\it WMAP} data.  These figures
also give an indication of constraints that will be available in the
future on reionization models -- by showing variations in each model
parameter that cause changes as small as $\Delta\tau=\pm0.01$ in the
electron scattering optical depth.  While the constraints on single
parameters are impressive at this $\Delta\tau$, considering one
parameter at a time will clearly be only the first crude step in
attempts to understand the reionization history.  Eventually, a full
degeneracy study, involving simultaneous variations of all
reionization model parameters, together with cosmological parameters,
will be needed.  While such a study is not yet called for by the {\it WMAP}
data, our results indicate that this will be {\it necessary} for Planck
(see paper II for in--depth discussion).

Such future studies will utilize information in the CMB polarization
and temperature power spectra beyond the value of the optical depth
$\tau$.  Indeed, the lack of our a--priori knowledge of the
reionization history would bias estimates of $\tau$ itself, and would
be the limiting factor in determining $\tau$ from Planck data (see
paper II).  Here we consider the question: {\it how useful is the
information beyond the measurement of $\tau$ in constraining
reionization histories?}

\begin{figure}[t]
\plotone{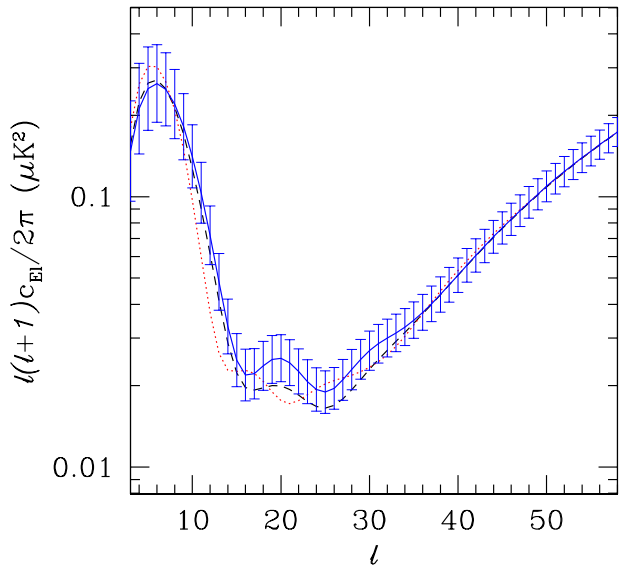}
\caption{The large--angle E--mode polarization anisotropies in three
different models. The dashed, dotted, and solid curves correspond to
the solid curves in Figures~\ref{fig:xe1}, \ref{fig:xe4feed}, and
\ref{fig:xe5}, respectively. All three models produce the same
electron scattering opacity $\tau=0.17$.  Cosmic variance error bars
are shown around the solid curve, in bins of $\Delta\ell=1$. The three
curves are statistically distinguishable in a cosmic variance limited
experiment at $>4\sigma$ significance. Planck can only distinguish the
dashed and the solid curves at $\sim 1.6\sigma$, but can distinguish
the other two pairs of models $>3\sigma$ (see Table~\ref{table:cmb}).}
\label{fig:ee} 
\end{figure}

\begin{figure}[t]
\plotone{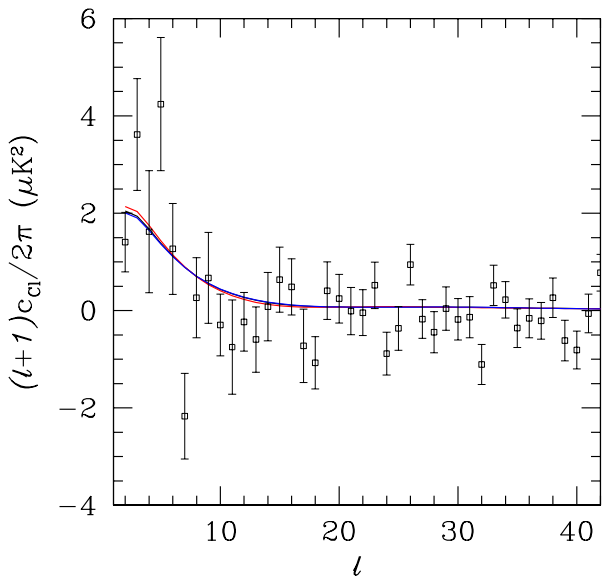}
\caption{The same three models as in Figure~\ref{fig:ee}, but the
temperature--polarization cross--correlation is shown.  The three
models are nearly indistinguishable.}
\label{fig:te} 
\end{figure}

To address this question, we here consider three different models that
have the same value of $\tau=0.17$.  Two of these models are the
fiducial models defined in \S~\ref{subsec:fiducial} and
\S~\ref{subsec:notypeII}, and shown by the solid curves in
Figures~\ref{fig:xe1} and \ref{fig:xe4feed}; hereafter referred to as
models (A) and (B).  The electron scattering in model (A) extends to
redshifts higher than in model (B), and, as a result, will produce
polarization anisotropy on smaller angular scales (larger spherical
harmonic index $\ell$).  For sake of illustration, we consider a third
model, similar to those proposed by Wyithe \& Loeb (2003) and Cen
(2003). This is a variant of model (A), in which we assume that a
transition from metal free to normal stellar populations takes place
abruptly at a fixed redshift $z_{\rm cr}$, accompanied by a sudden
decrease by a factor of $f_{\rm cr}$ in the ionizing photon production
rates. In terms of our model parameters,
$\epsilon_{II}=\epsilon_{Ia}=\epsilon_{Ib}=80$ at $z<z_{\rm cr}$ and
$\epsilon_{II}=\epsilon_{Ia}=\epsilon_{Ib}=80f_{\rm cr}$ at $z\geq
z_{\rm cr}$.  We find that $z_{\rm cr}=14$ and $f_{\rm cr}=20$ results
in $\tau=0.17$; this model is shown in Figure~\ref{fig:xe5} (together
with models with $z_{\rm cr}=10, 11, 13, 15, 17$, and $21$, which
bracket the $\Delta\tau\pm0.01$, $0.03$, $0.06$ range around
$\tau=0.17$).  Our motivation to reconsider this model, hereafter
referred to as model (C), is the distinctly different shape of the
reionization history.

We modified CMBFast\footnote{available at
http://www.physics.nyu.edu/matiasz/CMBFAST/online.html} (Seljak \&
Zaldarriaga 1996) to use the ionization histories from models (A), (B)
and (C) to generate temperature, polarization, and cross--correlation
anisotropy power spectra, $\ctl,\cel$ and $\ccl$ respectively.  The
resulting EE and TE power spectra are shown by the dashed, dotted, and
solid curves in Figures~\ref{fig:ee} and \ref{fig:te}, respectively.
In Figure~\ref{fig:ee}, we show cosmic variance error bars in bins of
$\Delta\ell=1$ around model (C), and in Figure~\ref{fig:te}, we
overlay the {\it WMAP} data.  The models are nearly indistinguishable
in the TE power spectra, but are clearly different in the EE spectrum:
model (A) is biased to lower $\ell$ than model (B), and model (C) has
excess power at small angular scales.

We next calculate the likelihood between pairs of models, using all
three (EE, TE, and TT) power spectra, and a $\Delta \bar{\chi}^2$
statistic defined in paper II (see also Kaplinghat et al. 2003).  As
apparent from Figures~\ref{fig:ee} and \ref{fig:te}, nearly all of the
distinguishing power comes from the E--mode polarization anisotropy
alone. In Table~\ref{table:cmb}, we show the significance at which
pairs of models can be distinguished in the future by (1) 2--year of
{\it WMAP} data (2) Planck, and (3) an ``ultimate'' cosmic variance limited
experiment.  The details of folding the instrumental characteristics
of {\it WMAP} and Planck into our calculations are described in paper II.

The three curves are statistically distinguishable in a cosmic
variance limited experiment at $>4\sigma$ significance. Planck can
only distinguish models (A) and (C) at $\sim 1.6\sigma$, but can
distinguish the other two pairs of models at $>3\sigma$ significance
(see Table~\ref{table:cmb}).  It is also interesting to ask how poor a
fit a ``one--step'' reionization model would be to real data realized
from a more complex history?  This issue is discussed in detail
in paper II; for sake of completeness, we quote here only the
difference of model (A) from one-step reionization models.
We find likelihoods of $\Delta \bar{\chi}^2$ of 0.3, 16, and 60, for
2--yr {\it WMAP} data, Planck, and a cosmic variance limited experiment,
respectively.  

In summary, our results suggest that for the physically motivated
reionization histories discussed in this paper, the Planck satellite
will be able to rule out both a simple 1--step reionization model, and
also several ``wrong'' physically motivated models, at high
statistical significance.

\begin{table*}[b]\small
\caption{\label{table:cmb} Constraints ($\Delta \bar{\chi}^2$) from
future CMB polarization experiments on three models with the same
$\tau=0.17$.}
\begin{center}
\begin{tabular}{cccc}
 Model pair & 2--yr {\it WMAP} & Planck & Cosmic variance \\
\hline
AB     & 0.2   &  9    & 25 \\
AC     & 0.1   &  2.6   & 18 \\
BC     & 0.3   &  13    & 34  \\
\hline
A vs one-step & 0.3  &   16 & 60\\
\multicolumn{4}{l}{} \\
\end{tabular}\\[12pt]
\end{center}
\end{table*}

\section{Can JWST detect the sources of reionization?}
\label{sec:jwst}

A broad conclusion that can be drawn from the {\it WMAP} results, reinforced
in this paper, is that UV sources exist at very high redshifts ($z\sim
20$).  This, by itself, is a significant new discovery, and it raises
the question of whether we will be able to directly detect these
sources.  Here we consider the capabilities of the {\it James Webb
Space Telescope (JWST)}: {\it what is the ``limiting redshift'' out to
which JWST will be able to detect the sources of reionization?}  We
here answer this question with a simple calculation, as follows.
Consider a halo of mass $M$ at redshift $z$.  The number of halos
$N(>M)$ with mass above $M$ at redshift $z\pm dz/2$ that could be
visible within a a solid angle $\Delta\Omega$ and redshift range
$\Delta z$ is given by \beq N(>M) = \Delta\Omega\times\Delta z\times
f_{\rm on}\times \left[ \frac{dV}{dzd\Omega}\left(z\right)
\int_{M}^\infty dM^\prime \frac{dn}{dM^\prime} \right],
\label{eq:dNdzdom}
\eeq where $dV/dzd\Omega$ is the cosmological volume element, 
$dn/dM $ is the halo mass function from Jenkins et al. (2001), and
$f_{\rm on}$ is the fraction of the time the halo spends at a
luminosity detectable by {\it JWST}.  We assume the sources live for
$t_s=3\times 10^6$ years (approximately the main sequence lifetime for
massive stars, both normal and metal--free, and a conservative
estimate for the bright phase of mini--quasar activity). In this case,
$f_{\rm on}\approx t_s/t(z)\approx 0.006-0.02$ at redshifts $z=10-20$
(where $t(z)$ is the age of the universe at redshift $z$).

The sensitivity expected to be reached in a $5\times 10^4$second
exposure ``{\it JWST} Deep Field'', with the current 6m diameter
telescope design, in the $0.6-5\mu$m band (S/N=5), with the Near
Infrared Camera (NIRCam) is $\sim 1$nJy.  We require that there should
be at least 100 sources visible down to this depth in the
$\Delta\Omega=10$ arcmin$^2$ field of view of NIRCam, and in $\Delta
z\approx 2$.  To predict the luminosities, consistent with our
reionization models, we make the assumption that in a halo of mass
$M$, $\sim 10\%$ of the gas turns into stars with a normal Salpeter
IMF.  We use the population synthesis models of Leitherer et
al. (1999) to predict the observed spectra (at a metallicity of
$Z=0.02$). We ignore wavelengths shorter than $1215(1+z)$\AA, because
of the strong GP trough expected at high redshifts, and compute the
average flux in the unobscured fraction of the $0.6-5\mu$m wavelength
range range.

Under these assumptions, we find that a halo of mass $2\times
10^8\,{\rm M_\odot}$ at redshift $z=14$ would have an average flux of
$\sim 1$nJy in the unobscured $1.8-5\mu$m range, detectable with
NIRCam.  Such a halo represents a $3\sigma$ object at $z=14$ in the
{\it WMAP} cosmology, and with our derived duty--cycle of $f_{\rm
on}=0.01$ there should be 100 such sources detectable at any given
time in $10$ arcmin$^2$.  This halo would have a virial temperature of
$\sim 2\times10^4$ K, justifying the assumed star formation efficiency
of $10\%$ (larger than is expected in Type II halos).  As we argued
above in \S~\ref{subsec:efficiencies}, it is possible that most stars
are still metal--free in Type Ia halos at these high
redshifts. Although metal free stars produce more ionizing photons
than normal stars, owing to their high effective temperatures, they
are fainter at the {\it JWST} wavelengths than normal stars. This,
however, is likely offset by the steep IMF expected from a metal--free
stellar population.  We conclude that {\it JWST} could directly detect
the sources of reionization, out to redshifts as high as $z\sim 14$;
these sources would be sufficiently bright to allow follow--up
spectroscopy with NIRSpec on {\it JWST}.  The sources could also be
discovered in line emission, either in the H$\beta$ line (Oh 1999),
or, in case of metal--free stellar populations, in the He 1640\AA, and
4686\AA\, lines out to redshifts $z\sim 20$ with the mid--infrared
instrument (Oh, Haiman \& Rees 2001; Tumlinson, Giroux \& Shull 2001).

\section{Conclusions}
\label{sec:discussion}

The determination by {\it WMAP} of a high optical depth to Thomson
scattering implies that significant reionization took place at
redshifts $z\sim 20$, and has provided our first glimpse into the
earliest structures that exist at these distant epochs.  In this
paper, we used physically motivated semi--analytical reionization
models to quantify the implications of the {\it WMAP} result.  The high
value of $\tau$, by itself, requires that the earliest light sources
have a surprisingly high efficiency $\epsilon$ of injecting ionizing
photons into the IGM. In the running spectral index cosmological model
favored by the combination of {\it WMAP} with other large scale structure
data (Spergel et al. 2003), ${\rm H_2}$ molecules are required to form
efficiently in halos with virial temperatures $T_{\rm vir}<10^4$K at
high redshifts to produce $\tau=0.17$.

Our most interesting conclusions arise when we require that our models
produce (1) a high optical depth $\tau=0.17$, and (2) a percolation
epoch at $z\sim 6-7$.  The latter is required by most interpretations
of the rapid evolution in the neutral fraction in the IGM at $z=5-6$
(Songaila \& Cowie 2002; Becker et al. 2001), and possibly by the high
temperature of the IGM inferred from the Ly$\alpha$ forest at $z\sim
4$ (Hui \& Haiman 2003). We generically find that no simple model can
simultaneously satisfy both of the above requirements.  Our results indicate
that either ${\rm H_2}$ molecules had to form efficiently at $z\sim
20$ and allow UV sources in halos with virial temperatures $T_{\rm
vir}<10^4$K to contribute substantially to reionization, or else the
ionizing photon production efficiency in halos with $T_{\rm
vir}>10^4$K had to decrease by a factor of $\gsim 30$ between $z\sim
20$ and $z\sim 6$.  The first of these two options would have
important implications on radiative feedback processes that determine
the global ${\rm H_2}$ chemistry.  The second option would naturally
be explained by the presence of metal--free stars in the earliest
halos, which are then replaced by normal stellar populations in larger
halos that collapse at later epochs.

The physics of reionization is rich in features that can naturally
lead to distinctive ionization histories.  These features can arise
because of (1) the different types of coolants in halos with virial
temperatures above and below $\sim10^4$K, (2) the different response
of different halos to radiative feedback on the ${\rm H_2}$ chemistry,
and to photoionization feedback on gas infall, and (3) the different
properties of metal--free and normal stellar populations.  A range of
physically motivated reionization histories is allowed by current
data. We find that the ionization history in our models generally
evolves monotonically with redshift, in difference from ``double
reionization'' that results if the metallicity was increasing
uniformly in the IGM, triggering a sharp transition at a fixed
redshift from metal--free to ``normal'' stars.  Nevertheless, we find
that the evolution of the ionized fractions have sufficiently
conspicuous features for Planck to provide tight statistical
constraints on reionization model parameters, and to elucidate much of
the physics at the end of the Dark Ages.  The sources responsible for
the high optical depth discovered by {\it WMAP} should be directly
detectable out to $z\sim 15$ by the {\it James Webb Space Telescope}.

The {\it WMAP} result has opened a new window into studies of the first
structures at the end of the cosmological dark ages, and has raised
the hopes for further interesting findings from all potential probes
of the high--redshift universe; including future CMB polarization
anisotropy studies, spectroscopic observations of quasars and
Ly$\alpha$ emitting galaxies at $z>6$, radio probes of the redshifted
21cm line of neutral hydrogen, as well as direct detections of the
sources of reionization and their end--products as supernovae and
gamma ray bursts.

\acknowledgements{We thank Manoj Kaplinghat, Lloyd Knox, and David
Spergel for stimulating discussions.  GPH is supported by the
W.M. Keck foundation. We thank Uros Seljak and Matias Zaldarriaga for
the use of their CMBFast code.}


\begin{thebibliography}{}

\bibitem{abn00} Abel, T., Bryan, G. L., \& Norman, M. L. 2000, ApJ, 540, 39

\bibitem{abn02} Abel, T., Bryan, G. L., \& Norman, M. L. 2002, Science, 295, 93

\bibitem{ah01} Abel, T., \& Haiman, Z. 2001, in ``Molecular hydrogen in space'',  Cambridge contemporary astrophysics, Eds. F. Combes, and G. Pineau des Forêts, Cambridge University Press: Cambridge, UK, p. 237


\bibitem{bl99} Barkana, R., \& Loeb, A. 1999, ApJ, 523, 54

\bibitem{bl01} Barkana, R., \& Loeb, A. 2001, Physics Reports, 349, 125

\bibitem[{Becker} {\em et~al.} (2001)]{becker01} {Becker}, R.~H. {\em et~al.} 2001, \aj, 122, 2850--2857.

\bibitem{wmap} Bennett, C. L., et al. 2003, ApJ, submitted, astro-ph/0302207 

\bibitem{bcl01} Bromm, V., Coppi, P. S., \& Larson, R. B. 1999, ApJ, 527, 5 

\bibitem{bcl02} Bromm, V., Coppi, P. S., \& Larson, R. B. 2002, ApJ, 564, 23

\bibitem{betal01} Bromm, V., Ferrara, A., Coppi, P. S., \& Larson, R. B. 2001, MNRAS, 328, 969

\bibitem{bkl01} Bromm, V., Kudritzki, R. P., \& Loeb, A. 2001, ApJ, 552, 464

\bibitem{bl02} Bromm, V., \& Loeb, A. 2002, ApJ, submitted, astro-ph/0212400

\bibitem{co00} Chiu, W. A., \& Ostriker, J. P. 2000, ApJ, 534, 507

\bibitem{cf00} Ciardi, B., Ferrara, A., \& Abel, T. 2000, ApJ, 533, 594

\bibitem[{Cen} (2003)]{cen03} {Cen}, R. 2003, \apj, submitted, astro--ph/0210473.

\bibitem[{Cen} and {McDonald} (2002)]{cen02} {Cen}, R. and {McDonald}, P. 2002, \apj, 570, 457--462.

\bibitem{dsf00} Dove, J. B., Shull, J. M., \& Ferrara, A. 2000, ApJ, 531, 846

\bibitem[Eisenstein and Hu (1999)]{eh99} Eisenstein, D. J., \& Hu, W. 1999, ApJ, 511, 5

\bibitem[Efstathiou(1992)]{efstathiou} Efstathiou, G. 1992, MNRAS, 256,
43

\bibitem{fetal00} Fan, X., et al. 2000, AJ, 120, 1167

\bibitem{fetal03} Fan, X., et al. 2003, AJ, in press, astro-ph/0301135

\bibitem[{Fan} {\em et~al.} (2002)]{fan02} {Fan}, X., {Narayanan}, V.~K., {Strauss}, M.~A., {White}, R.~L., {Becker},   R.~H., {Pentericci}, L., and {Rix}, H. 2002, \aj, 123, 1247--1257.

\bibitem{f98} Ferrara, A. 1998, ApJ, 499, L17

\bibitem{fhp98} Fukugita, M., Hogan, C. J., \& Peebles, P. J. E. 1998, ApJ, 503, 518

\bibitem{gb02} Glover, S. C. O., \& Brandt, P. W. J. L. 2003, MNRAS, in press, astro-ph/0205308

\bibitem{g01} Gnedin, N.Y. 2001, MNRAS, submitted, astro-ph/0110290

\bibitem{go97} Gnedin, N. Y., \& Ostriker, J. P. 1997, ApJ, 486, 581

\bibitem[Gnedin and Shandarin (2002)]{gs02} Gnedin, N. Y., \& Shandarin, S. F. 2002, MNRAS, 337, 1435

\bibitem{gp65} Gunn, J. E., \& Peterson, B. A., 1965, ApJ, 142, 1633

\bibitem{h03} Haiman, Z. 2003, Carnegie Observatories Astrophysics Series, Vol. 1: Coevolution of Black Holes and Galaxies, ed. L. C. Ho (Cambridge: Cambridge Univ. Press), submitted

\bibitem{ham01} Haiman, Z., Abel, T., \& Madau, P. 2001, ApJ, 551, 599

\bibitem[{Haiman}, {Abel}, and {Rees} (2000)]{har00} Haiman, Z., Abel, T., \& Rees, M. J. 2000, ApJ, 534, 11

\bibitem[Haiman and Loeb (1997)]{hl97} Haiman, Z., \& Loeb, A. 1997, ApJ, 483, 21

\bibitem{hl98} Haiman, Z., \& Loeb, A. 1998, ApJ, 503, 505

\bibitem{hc99} Haiman, Z., \& Knox, L. 1999, in ``Microwave Foregrounds'', eds. A. de Oliveira-Costa \& M. Tegmark (San Francisco: ASP), p. 227

\bibitem[{Haiman}, {Rees}, and {Loeb} (1997)]{hrl97} Haiman, Z., Rees, M. J., \& Loeb, A. 1997, ApJ, 476, 458 [erratum: 1997, ApJ, 484, 985]

\bibitem{htl96} Haiman, Z., Thoul, A. A., \& Loeb, A. 1996, ApJ, 464, 523

\bibitem{papII} Holder, G. P., Haiman, Z., Kaplinghat, M., \& Knox, L. 2003, ApJL, submitted, astro-ph/0302404 (paper II)

\bibitem{hg97} Hui, L., \& Gnedin, N. Y. 1997, MNRAS, 292, 27

\bibitem{hh03} Hui, L., \& Haiman, Z. 2003, to be submitted to ApJ

\bibitem[{Jenkins} {\em et~al.} (2001)]{jenkins01} {Jenkins}, A., {Frenk}, C.~S., {White}, S.~D.~M., {Colberg}, J.~M., {Cole}, S.,   {Evrard}, A.~E., {Couchman}, H.~M.~P., and {Yoshida}, N. 2001, \mnras, 321, 372.

\bibitem[{Kaplinghat} {\em et~al.} (2003)]{kaplinghat03} {Kaplinghat}, M., {Chu}, M., {Haiman}, Z., {Holder}, G.~P., {Knox}, L., and   {Skordis}, C. 2003, \apj, 583, 24--32.

\bibitem[Kogut {\em et~al.} (2003)]{kogut03} Kogut, A. 2003, ApJ,  submitted, astro-ph/0302213

\bibitem{l99} Leitherer, C., Schaerer, D., Goldader, J. D., Delgado, R. M. G., Robert, C., Kune, D. F., de Mello, D. F., Devost, D., Heckman, T. M. 1999, ApJS, 123, 3; electronic data are available at http://www.stsci.edu/science/starburst99/

\bibitem{letal02} Lidz, A., Hui, L., Zaldarriaga, M., \& Scoccimarro, R. 2002, ApJ, 579, 491

\bibitem[{Loeb} and {Barkana} (2001)]{barkana01} {Loeb}, A. and {Barkana}, R. 2001, \araa, 39, 19--66.

\bibitem{mba01} Machacek, M. E., Bryan, G. L., \& Abel, T. 2001, ApJ, 548, 509

\bibitem{mba03} Machacek, M. E., Bryan, G. L., \& Abel, T. 2003, MNRAS, 338, 273

\bibitem{mhr99} Madau, P., Haardt, F., \& Rees, M. J. 1999, ApJ, 514, 648

\bibitem{mf99} MacLow, M.-M., \& Ferrara, A. 1999, ApJ, 513, 142

\bibitem{mhr00} Miralda-Escud\'e, J., Haehnelt, M., \& Rees, M. J. 2000 ApJ, 530, 1

\bibitem{jordi01} Miralda-Escud\'e, J. 2001, in ``The Physics of Galaxy Formation'', ASP Conference Proceedings, Vol. 222, eds. M. Umemura and H. Susa. (San Francisco: ASP), p.181

\bibitem[Navarro \& Steinmetz(1997)]{navarro} Navarro, J. F., \& Steinmetz, M. 1997, ApJ, 478, 13

\bibitem{peng99} Oh, S. P. 1999, ApJ, 527, 16

\bibitem{ohr01} Oh, S. P., Haiman, Z., \& Rees 2001, ApJ, 553, 73

\bibitem{on99} Omukai, K., \& Nishi, R. 1999, ApJ, 518, 64

\bibitem{petal02} Pentericci, L., et al. 2002, AJ, 123, 2151

\bibitem{retal02} Razoumov, A. O., Norman, M. L., Abel, T., \& Scott, D. 2002, ApJ, 572, 695

\bibitem{rgs02} Ricotti, M., Gnedin, N. Y., \& Shull, J. M. 2002, ApJ, 575, 49

\bibitem[{Seljak} and {Zaldarriaga} (1996)]{seljak96} {Seljak}, U. and {Zaldarriaga}, M. 1996, \apj, 469, 437

\bibitem{schaerer02} Schaerer, D. 2002, A\&A, 382, 28

\bibitem[Shapiro and Giroux (19870]{sg87} Shapiro, P., \& Giroux, M.L. 1987, ApJ, 321, L107

\bibitem{sbg94} Shapiro, P. R., Giroux, M. L., \& Babul, A. 1994, ApJ, 427, 25

\bibitem{srm98} Shapiro, P. R., Raga, A. C., \& Mellema, G. 1998, in Molecular Hydrogen in the Early Universe, Memorie Della Societa Astronomica Italiana, Vol. 69, ed. E. Corbelli, D. Galli, and F. Palla (Florence: Soc. Ast. Italiana), p. 463

\bibitem{sc02} Songaila, A., \& Cowie, L. L. 2002, AJ, 123, 2183

\bibitem[{Spergel} et al. (2003)]{spergel03} Spergel, D. N. et al. 2003, ApJ, submitted, astro-ph/0302209

\bibitem{spa01} Steidel, C. C., Pettini, M., \& Adelberger, K. L. 2001, 546, 665

\bibitem{tetal94} Tegmark, M., Silk, J., \& Blanchard 1994, ApJ, 420, 484

\bibitem{tetal97} Tegmark, M., Silk, J., Rees, M. J., Blanchard, A., Abel, T., \& Palla, F. 1997, ApJ, 474, 1

\bibitem{ts01} Tumlinson, J., Giroux, M. L., \& Shull, J. M. 2001, ApJ, 551, 1

\bibitem{tgs01} Tumlinson, J., \& Shull, J. M. 2000, ApJ, 528, 65

\bibitem[Thoul \& Weinberg(1995)]{tw95} Thoul, A. A., \& Weinberg, D. H. 1995, ApJ, 442, 480

\bibitem{vs99} Valageas, P., \& Silk, J. 1999, A \& A, 347, 1

\bibitem[{{Venkatesan}(2000)}]{venkatesan00} {Venkatesan}, A. 2000, \apj, 537, 55

\bibitem[{{Venkatesan}(2002)}]{venkatesan02} Venkatesan, A. 2002, \apj, 572, 15

\bibitem{wl00} Wood, K. \& Loeb, A. 1999, ApJ, 545, 86

\bibitem[{Wyithe} and {Loeb} (2003)]{wyithe03} {Wyithe}, S. and {Loeb}, A. 2003, \apj, in press, astro--ph/0209056.

\bibitem{z97} Zaldarriaga, M. 1997, Phys. Rev. D, 55, 1822

\end{thebibliography}
\end{document}